\documentclass[12pt]{article}
\usepackage{graphicx,color,amssymb,amsmath}
\begin{document}
\title{\textbf{$t'$ at the LHC: the physics of discovery}} 
\author{B. Holdom%
\thanks{bob.holdom@utoronto.ca}\\
\emph{\small Department of Physics, University of Toronto}\\[-1ex]
\emph{\small Toronto ON Canada M5S1A7}}
\date{}
\maketitle
\begin{abstract}
A search for a fourth family at the LHC is presently a low priority, but we argue that an effective search can be conducted early with only a few inverse femtobarns of data. We discuss a method based on invariant masses of single jets for identifying the $W$'s originating from heavy quark decays. This can significantly increase signal to background in the reconstruction of the $t'$ mass. We also study the various types of physics that can impact the background estimate, most notably higher order effects, initial state radiation, and models of the underlying event.
\end{abstract}

\section{Introduction}
There are sometimes speculations about a worst case scenario unfolding at the LHC. For example a light Higgs finally showing up after many years of running with no further signs of new physics would be distinctly unsatisfying. This would provide us with no insight about the origins of the observed mass spectrum and the flavor structure of the standard model. The possibility of not finding \textit{anything}, including the Higgs, is equally depressing since it leaves open the possibility that new physics lurks just beyond the reach of the LHC. In contrast, a best case scenario is one that maximizes our understanding of electroweak symmetry breaking, as well as shedding light on the questions of mass and flavor. Maximizing understanding is closely related to a maximal elimination of the widest range of possibilities currently being considered by theorists.\footnote{Experimentalists, at least, may refer to this as a best case scenario.} This would not be accomplished very quickly for example if the first anomalies involved missing energy, due to escaping particles stabilized by a new conserved quantum number. A wide range of different theories may have similar missing energy signatures, and it could take considerably more effort to pin down the actual physics.

Here we argue that a best case scenario involves the discovery of a sequential fourth family (sequential means quarks and leptons with standard quantum numbers). If the masses of these new fermions lie in a certain range, then it implies that we have identified the dominant order parameter of electroweak symmetry breaking. Strong interactions are implied, and the whole idea of a perturbative description of electroweak symmetry breaking, as embodied in models of light elementary Higgs scalars, low energy supersymmetry, and composite Higgs constructions, is eliminated. In addition, as we argue in this note, there may be signals for this possibility that can show up quite early, thus making this at least or more exciting than other possible strong interaction scenarios such as technicolor and related dual descriptions in higher dimensions.

Our scenario differs in a fundamental respect from these other strong interaction scenarios. Consider the nature of the propagating degrees of freedom most closely associated with electroweak symmetry breaking. In other words what are the propagating degrees of freedom to which the Goldstone bosons couple most strongly, and among these which are the lightest? These degrees of freedom are typically bosonic. Typically some rho-like resonance plays this role, or some Kaluza-Klein mode, or some scalar Higgs-type mode. These bosons are often produced singly as resonances in colliders. In our scenario the prime signal will instead involve the pair production of unconfined fermions, whose masses are the order parameters for electroweak symmetry breaking. Each fermion decays weakly, which for the case of decaying quarks produces the $f\overline{f}W^{+}W^{-}$ signal where $f$ is typically $b$ or $t$. A bosonic resonance in contrast typically decays to a fermion pair \textit{or} a pair of gauge bosons. In this note our interest is in the $b\overline{b}W^{+}W^{-}$ signal.

A sequential fourth family is not the unique possibility for new heavy fermions. For example new fermions that are vector-like under the electroweak symmetries are allowed to have a wide range of masses unrelated to electroweak symmetry breaking. In this sense their discovery would be less informative than a sequential fourth family. There are two ways to distinguish a sequential fourth family from vector-like fermions and other exotic possibilities. One is that the fourth family masses would have to satisfy the constraints arising from electroweak precision measurements. In particular the mass splittings of the quark and lepton doublets and the relative sizes of the quark and lepton masses are constrained  \cite{O}, and if such mass relations are observed then this would constitute good evidence.\footnote{Of course the fourth neutrino must not be light, lying at least above $\sim 80$ GeV, but remember that the fourth family leptons are also associated with strong interactions.} The other is that our proposed signal involves the weak charged-current decay of new heavy fermions. In contrast new vector-like fermions may have dominant decay modes through gluon or $Z$ emission. The suppression of these flavor-changing neutral currents is natural for sequential fermions, but less so for nonstandard fermions that can mix with standard fermions in a variety of ways.

Our study will focus on the weak decay of a sequential, charge $2/3$, quark, the $t'$, assumed to have mass in the 600-800 GeV range. This range should be typical of a dynamically generated fermion mass if this is the mass of electroweak symmetry breaking, assuming no fine tuning in the underlying dynamics.  Related to this is the old observation that 550 GeV is roughly the mass of a heavy quark above which its coupling to the Goldstone boson is strong \cite{N}. For more discussion of the dynamics and structure of such a theory see \cite{O}. For most of our study we use $m_{t'}=600$ GeV, but we also briefly compare to $m_{t'}=800$ GeV. We assume that the fourth family enjoys CKM mixing and that the resulting decay $t'\rightarrow bW$ is dominant. This is naturally the case if $m_{b'}>m_{t'}$,\footnote{An attempt to understand the top mass in the context of a fourth family yields $m_{b'}>m_{t'}$ \cite{O}.} and it is still true for a range of the mixing element $V_{t'b}$ when $m_{t'}-M_W<m_{b'}<m_{t'}$ so that $t'\rightarrow b'W^{(*)}$ is suppressed by phase space.

With the $t'\rightarrow bW$ decay dominant then the width of the $t'$ is roughly 60 $|V_{t'b}|^2$ GeV, and thus $t'$ could well be narrower than $t$ for $V_{t'b}$ reasonably small. We are not considering the possibility of single $t'$ production here, but that cross section is also proportional to $|V_{t'b}|^2$. If $|V_{t'b}|\lesssim 0.1$ for example then $t'\overline{t'}$ production should dominate \cite{R}.

$t'\overline{t}'$ production adds to the same final states as $t\overline{t}$ production, and thus a reconstruction of the $t'$ mass is likely necessary to pull the signal from background.  In the analyses done thus far a strategy similar to the original $t$ mass determination is adopted; the resulting signal to background ratio $S/B$ is not very encouraging, making the search for a 600-800 GeV $t'$ fairly challenging even for 100 fb$^{-1}$ of data \cite{K}. A lower limit on the $t'$ mass using a similar analysis has been set by the CDF collaboration \cite{S}.

Here we present a preliminary study of a method that appears to significantly improve signal to background $S/B$. We make use of the fact that standard jet reconstruction algorithms may tend to combine the two proto-jets from the hadronic decay of a sufficiently energetic $W$ into a single jet. Such a jet can be identified through a measurement of its invariant mass, which can be obtained from the energy deposits in the calorimeter cells without any need to resolve nearly merged proto-jets. The idea of using jet masses to identify highly energetic $W$'s (and as well $t$'s) has been considered before in efforts to reconstruct the $t$ mass in $t\overline{t}$ production \cite{J,F}. Our motivation here is in a sense orthogonal in that the reconstruction of $W$ jets in this way may actually act to suppress the $t\overline{t}$ background relative to signal. The reason is basically kinematical, having to do with the relative isolation of $W$ jets and $b$ jets in signal versus background. We should stress that $W$'s are being identified through their hadronic rather than leptonic decay modes and this allows us to consider event selection without requirements for leptons or missing energy. This increases statistics considerably.

\section{Event selection}
We utilize the PGS4 detector simulation program \cite{Y} which conveniently reconstructs the invariant masses of jets. PGS4 incorporates both a cone-based jet finding algorithm and a $k_T$-based algorithm. See \cite{Z} for a description and comparison of these algorithms as applied to the proto-jets of $W$ decay. We will mostly confine ourselves to the cone-based finder, and we show by comparison that it tends to produce a significantly better $S/B$ for our application. We use the ATLAS LHC detector simulation parameter choices,\footnote{These settings are found in the Madgraph package which includes PGS4.} which includes a $\Delta \eta\times\Delta\phi=0.1\times0.1$ grid size to model the hadronic calorimeters along with some estimate of their energy resolution. Our only change will be to the cone size for the cone-based jet finder, for which we choose 0.6. Since our own event selection is very restrictive we will not make use of triggers in the detector simulation.

The dominant irreducible background is $t\overline{t}$ production, whose cross section is $\sim 500$ times larger than for $t'\overline{t}'$. An obvious reduction of this background comes by imposing a lower bound $\Lambda_{\rm top5}$ on the scalar $p_T$ sum of some number (say five) of the hardest reconstructed objects in the detector. These objects may include leptons and missing energy. We will choose $\Lambda_{\rm top5}=2m_{t'}$. Because of the $\Lambda_{\rm top5}$ cut we can apply a lower bound $\Lambda_{\rm totE}$ on the scalar $p_T$ sum of all final particles in the Herwig or Pythia output, so as to reduce the number of events to be stored and/or simulated by PGS4. When $m_{t'}=600$ GeV and $\Lambda_{\rm top5}=1200$ GeV we choose $\Lambda_{\rm totE}=1000$ GeV. The latter cut, for computational purposes only, removes an insignificant fraction of the events that pass the other cuts. 

Tagging $b$-jets will play an important role in our analysis, and we will insist on at least one $b$-tag with $p_T>\Lambda_b$. We find that a good choice is $\Lambda_b=m_{t'}/3$. In an analysis of real data where $m_{t'}$ is not known, our $\Lambda_{\rm top5}$ and $\Lambda_b$ cuts will have to be varied to optimize the signal while keeping the ratio $\Lambda_{\rm top5}/\Lambda_b\approx 6$.

The $b$-tagging efficiencies incorporated in PGS4 have been fit to CDF data and are not very appropriate for our studies involving such energetic $b$-jets. With these energies the $b$-mistag rate involving light quarks and gluons is expected to deteriorate (increase) \cite{I}. To account for this and to make the dependence on the efficiencies more transparent we replace the $b$-tag efficiencies, both the tight and loose sets in PGS4, with a single set (1/2, 1/10, 1/30) for underlying $b$'s, $c$'s, and light quarks or gluons, respectively. We assume these efficiencies vanish for pseudorapidity $|\eta|>2$, thus roughly modeling the point in pseudorapidity at which efficiencies start to deteriorate \cite{I}. When we consider the $W+\rm{jets}$ background, it will basically be the $b$-mistags of gluon jets that determine its level.  Thus it is the ratio of the first and last of the three efficiencies above that is most relevant for $S/B$, and we believe that our choice is conservative.

We will also require evidence for at least one $W$ in a manner that we now describe. The object is to make use of the observation that $t'\overline{t}'$ production generates jets originating from energetic $W$'s and $b$'s that are quite isolated from each other. This is in contrast to the $W$'s and $b$'s in the $t\overline{t}$ background passing the above cuts, since they typically come from decays of quite highly boosted $t$'s. In addition, as we have mentioned, when the $W$'s are sufficiently energetic the two jets from the hadronically decaying $W$ will often be reconstructed as a single jet. In particular, even for $E_T=150$ GeV $W$ jets \cite{J}: ``Over 95\% of the jet energy is contained in a $\sqrt{{\Delta \eta}^2+{\Delta \phi}^2}=.7$ cone around the $E_T$ weighted baricenter." This then provides an opportunity to identify isolated and energetic $W$'s, by reconstructing jets using a $\Delta R\lesssim0.7$ cone (in fact we will use $\Delta R=0.6$), and then looking for a peak in the invariant mass distribution of these jets. The main point is that this procedure should be less efficient for the $t\overline{t}$ background events. The $W$ jets in these events will often be contaminated by the nearby $b$ jets, resulting in measured invariant masses that are more widely scattered relative to the true $W$ mass.

Note that our goal here is opposite to the usual and more complex task of trying to reconstruct $W$'s in $t\overline{t}$ events. In that case choosing a small cone size and using sophisticated analyses to reduce the merging and cross-contamination of jets is appropriate. Here we want these same effects to reduce the identification of $W$'s in the background sample, and a simple-minded cone algorithm with a fairly large cone size may be quite desirable for this purpose.

We are thus prompted to define a $W$-jet as a non-$b$-tagged jet whose invariant mass is close to a peak in the invariant mass distribution, which in turn is close to the $W$ mass. The actual peak location will be influenced by the effects of ``splash-out'' and ``splash-in''; the former occurs when the cone does not capture all of the energy originating from the $W$ decay (large angle contributions with respect to the cone center are weighted more in the invariant mass determination) and the latter occurs where the cone is receiving unrelated energy contributions from the underlying event and pile-up effects. There may also be scattering effects that spread the transverse shape of the jet that may not be properly modeled by the detector simulation. But the point is that both the peak location and width can be experimentally determined, and then a jet with an invariant mass falling appropriately close to the peak can be called a $W$-jet.

From the histograms for the jet invariant masses that we display below we are led to define a $W$-jet to be one whose invariant mass is within $\approx 10$ GeV of the peak. In fact we use a slightly optimized 9 GeV value. We require at least one $W$-jet defined in this way. Although our event selection relies on hadronic decays of the $W$, in our event generation we allow the $W$ to decay both hadronically and leptonically in both signal and background. In the sample of events containing both $b$ and $W$ jets we can now attempt a reconstruction of the $t'$ mass by considering the invariant mass of $b$-$W$ jet pairs. We consider all such pairs in each event, as long as the $b$-jet has $p_T>\Lambda_b$. We are thus interested in a pair of plots; one for the single jet invariant masses to display the $W$ peak, and one for the invariant masses of the $W$-$b$ pairs to display the $t'$ peak. Note that both plots are  produced from events that pass the $\Lambda_{\rm top5}$ and $\Lambda_b$ cuts. On each plot we will overlay signal and background histograms. 

We will produce these plots for each of a variety of the different event generation tools available: MC@NLO\cite{E}-Herwig, Herwig\cite{B}, Alpgen\cite{C}-Herwig, Alpgen-Pythia, Pythia\cite{A} and Madgraph\cite{D}-Pythia. We have made an effort to use up-to-date versions.\footnote{Herwig 6.51, MC@NLO 3.3, Alpgen 2.11, Madgraph 4.1, Pythia 6.409, PGS4 release 070120.} Our goal here is to study the effects of different types of physics that are modeled in various ways by these programs.
\begin{itemize}
\item MC@NLO comes the closest to correctly modeling the partonic scattering by including the one-loop effects.
\item For the study of more jets originating in the higher order tree level processes at the partonic level we use Alpgen, interfaced respectively to both Herwig and Pythia.\footnote{Note that some matching to massive quark matrix elements is already incorporated into Herwig and Pythia \cite{V}.}
\item Other physics having an important impact on the background is initial state radiation and the underlying event. Stand-alone Pythia includes more varied and possibly more advanced descriptions of this physics.
\item Different tools may be more convenient for different backgrounds; we shall use Madgraph-Pythia to model the $W+\rm{jets}$ background.
\item Pythia and Madgraph-Pythia make it easy to incorporate a fourth family and thus allow a parameterization of the CKM mixing relevant to $t'$ decay.
\end{itemize}

An important feature of our analysis is that for each case, we always use the same tool(s) to calculate both signal and background. In the next two sections we concentrate on the $t\overline{t}$ background and then after consider the $W+\rm{jets}$ background. We collect roughly 3 fb$^{-1}$ of integrated luminosity for the backgrounds. For the signal we often collect 5-10 times more and then scale the resulting histograms down; for both signal and background we scale results to 2.5 fb$^{-1}$. Although the signal histograms are then artificially smooth, they usefully show the structure expected as more data is collected.

\section{Event generation involving Herwig}
We first investigate the MC@NLO-Herwig combination, where MC@NLO corrects the partonic production process at the one-loop level, thus incorporating more correctly the first extra hard parton. It could be expected that the predicted size of both signal and background in MC@NLO is more reliable than the other approaches we consider below. (Other approaches require more consideration of renormalization scales and K-factors.) A slight drawback of MC@NLO is that it offers no straightforward way to incorporate a fourth family. Thus to model the process $t'\rightarrow bW$ we simply increase the $t$ mass to 600 GeV and then use $t\rightarrow bW$ to model $t'\rightarrow bW$. The larger than expected width of the $t'$ that this entails has little effect on our results. We also note that MC@NLO operates by producing a fraction ($\sim 15$\%) of events with negative weight, which need to be subtracted when forming histograms. We use the CTEQ6M parton distribution function (a NLO PDF) for use with MC@NLO-Herwig, but we also use it elsewhere (unless otherwise specified) to make comparisons more transparent. We note that the choice of PDF has minor impact on $S/B$ as long as the same PDF is used for both signal and background.
\begin{center}\includegraphics[scale=0.28]{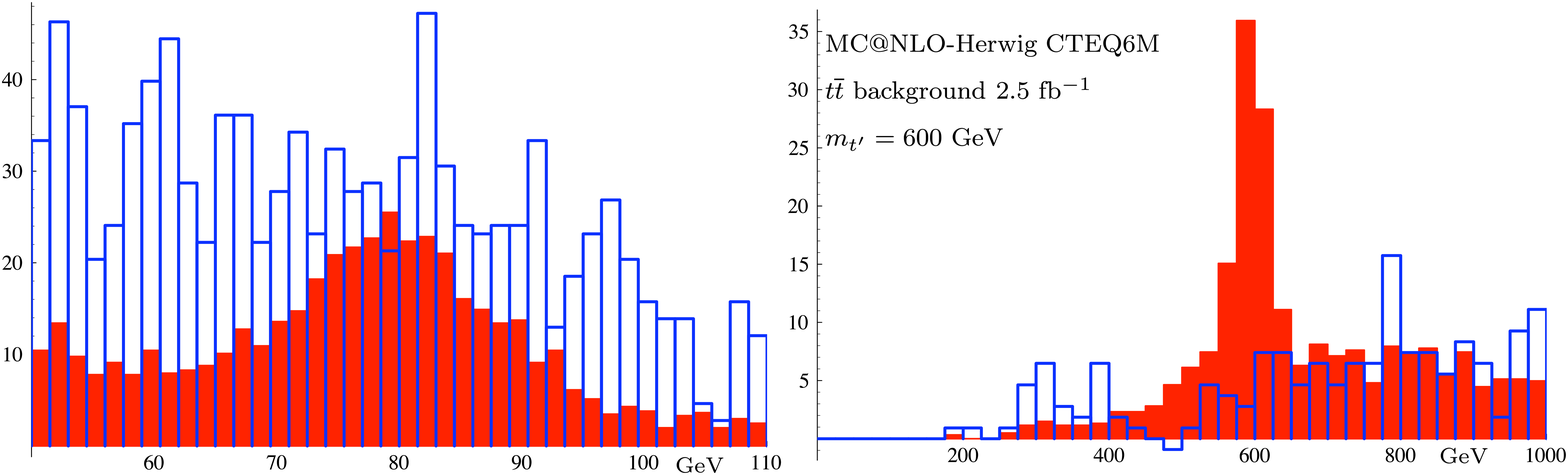}\end{center}
\vspace{-1ex}\noindent Figure 1: Signal (red) versus $t\overline{t}$ background (blue), using MC@NLO-Herwig. As for all such figures to follow, the $W$ mass plot is on the left and the $t'$ mass plot is on the right.
\vspace{2ex}

We present the resulting $W$ mass and $t'$ mass plots in Fig.~(1), comparing signal against the $t\overline{t}$ background. In the $W$ mass plot we see that a stronger peak at the $W$ mass shows up in the signal events as compared to the background events, as expected from our previous discussion, and which thus leads to an increased $S/B$. Even so we are surprised by how strong the $t'$ mass peak is relative to background. 

This naturally leads to the question of the role that MC@NLO is playing, and so we compare to Herwig when run in stand-alone mode so as to produce results without NLO corrections (the MC@NLO scripts provide this option). Those results are displayed in Fig.~(2). The comparison of these two sets of results show that the effect of MC@NLO is apparently to cause the signal to increase and the background to decrease! We also see that the large difference in the backgrounds in the $W$ mass plots do not carry over to the same degree in the $t'$ mass plots. This is at least partly due to the $\Lambda_{\rm top5}$ and $\Lambda_b$ cuts which, for the background, pushes a broad peak in the $W$-$b$ invariant mass spectrum to higher energies.
\begin{center}\includegraphics[scale=0.28]{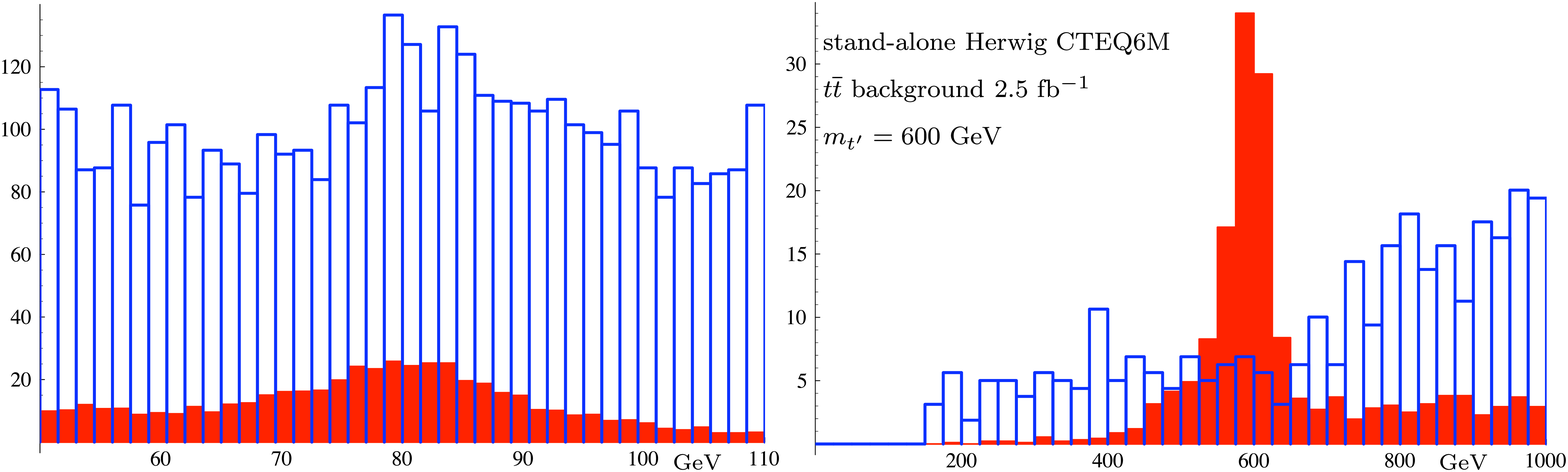}\end{center}
\vspace{-1ex}\noindent Figure 2: Signal versus $t\overline{t}$ background, using stand-alone Herwig.
\vspace{2ex}

For a better understanding of the large difference in background on the $W$ mass plots we consider the $H_T$ distribution, where $H_T$ is defined as the scalar sum of all transverse (including missing) momenta. Of interest is the high energy tail of this distribution for the $t\overline{t}$ background, since this is the region populated by the signal events. The $H_T$ distributions with and without MC@NLO are shown in Fig.~(3). On the high $H_T$ tail we see a significant reduction due to MC@NLO, even though MC@NLO increases the total cross section for $t\overline{t}$ production from 490 to 815 pb. In fact this increase in the total $t\overline{t}$ production cross section is similar to the increase in the $t'\overline{t'}$ production cross section, an increase from 0.87 to 1.33 pb. These increases are the K-factors. The K-factor for the signal combined with the change in shape of the $H_T$ distribution for the background gives some understanding of the signal enhancement and the background suppression. Also we see how tiny the signal appears in Fig.~(3); this highlights again the surprising effectiveness of the event selection and the $t'$ mass reconstruction in pulling the signal from the background.

Although we are finding that the physics incorporated by MC@NLO can have a significant effect on the shape of the $H_T$ distribution, we should keep in mind that the stand-alone Herwig results may be sensitive to choices made in its own attempt to model the physics. We have also considered the effect of the Jimmy add-on to Herwig to model the underlying event. We use a Jimmy tuning for the LHC, in particular \verb$PTJIM=4.9$ and \verb$JMRAD(73)=1.8$ \cite{M}. The result is more low energy activity in the event, but we find that this has only a minor effect on signal and background; the $H_T$ tail of the background is little affected. We have therefore presented results without Jimmy.
\begin{center}\includegraphics[%
  scale=0.56]{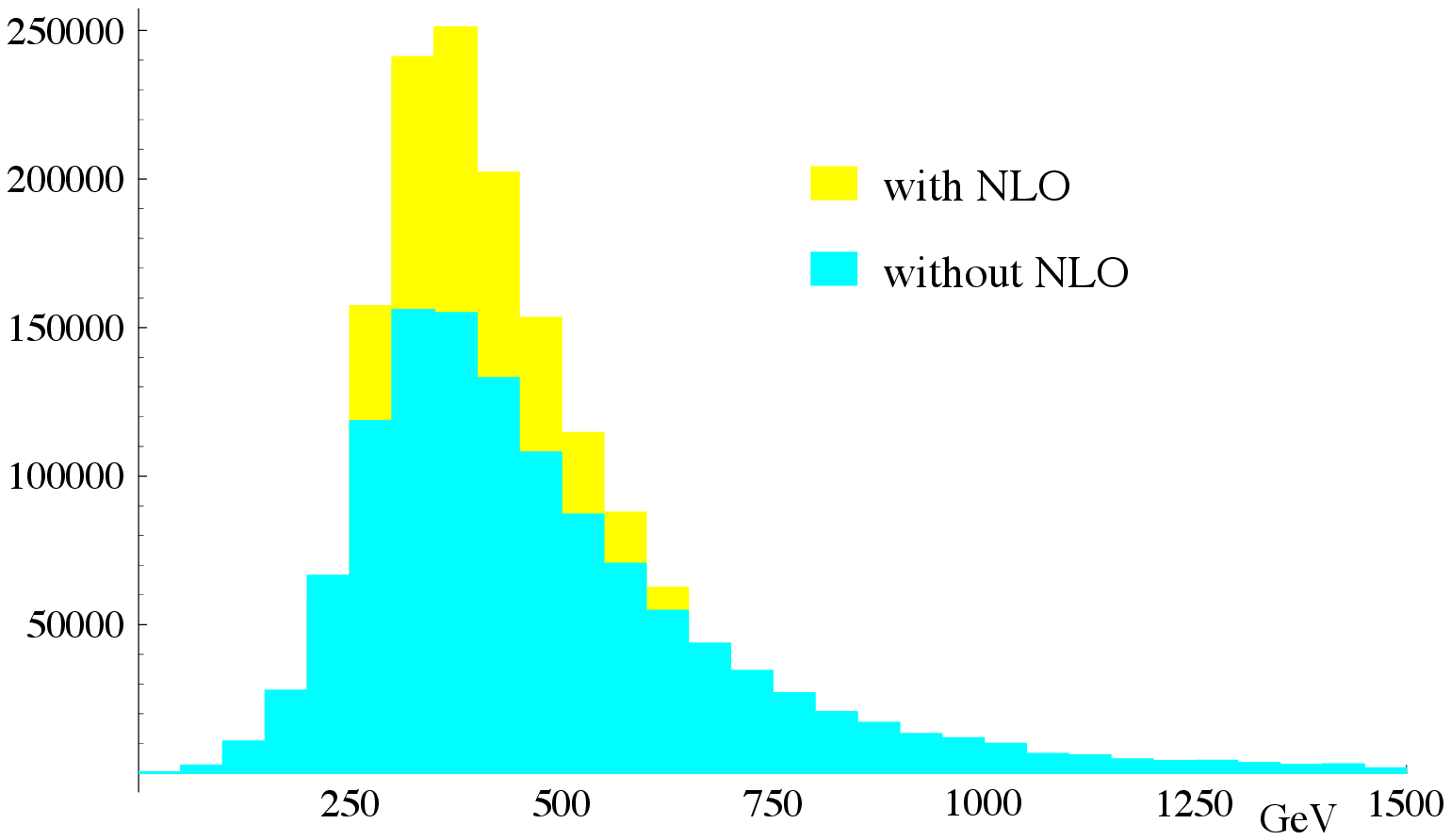}\hspace{-3pt}\includegraphics[%
  scale=0.56]{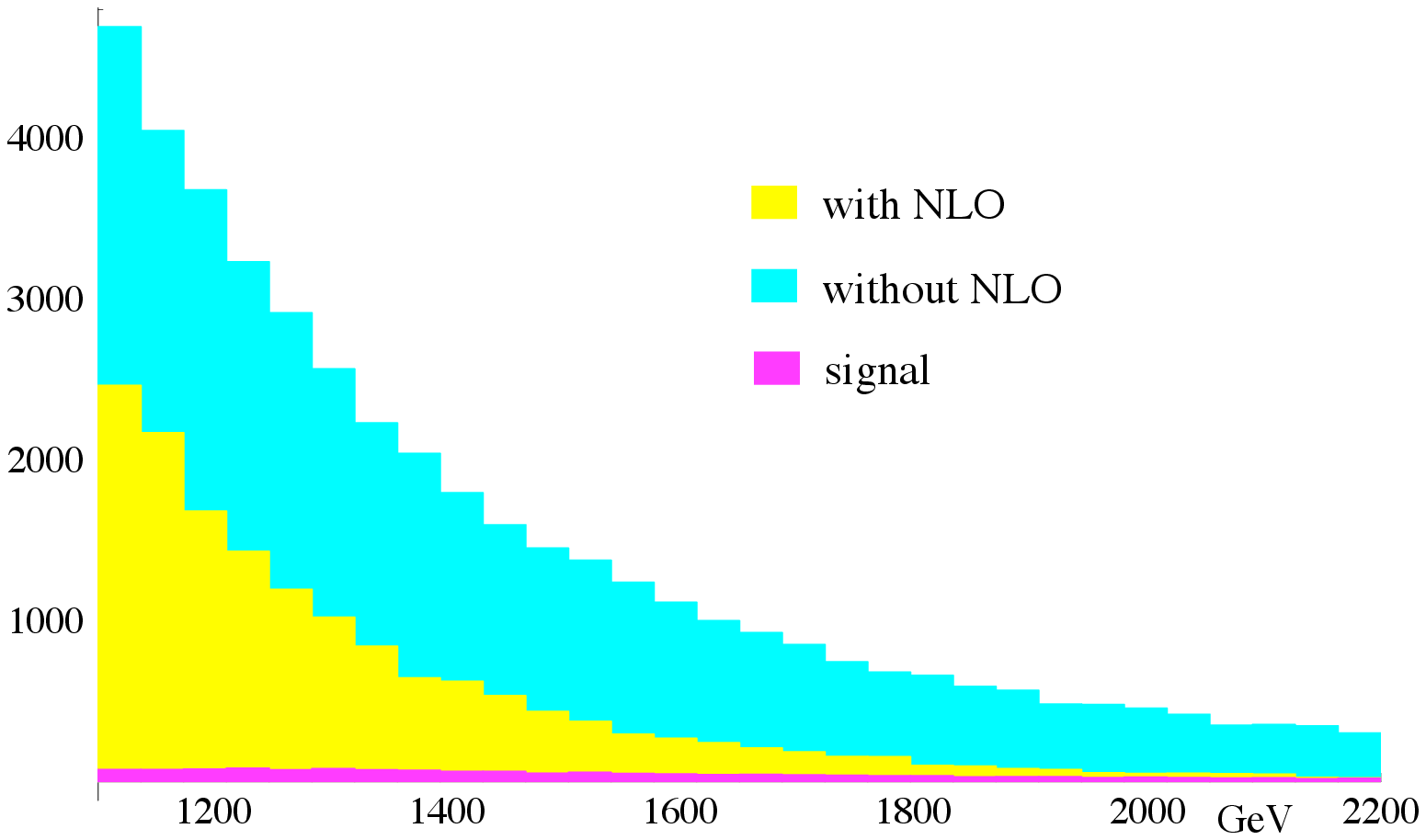}\end{center}
\vspace{-2ex}\noindent Figure 3: $H_T$ distributions for $t\overline{t}$ production, with and without NLO effects. The high $H_T$ tails are shown on the right along with the $t'\overline{t'}$ signal distribution.
\vspace{2ex}

Complimentary to MC@NLO in the effort to go beyond lowest order matrix elements is Alpgen, which can also interface with Herwig. Alpgen incorporates higher order amplitudes involving more partons and uses jet-parton matching to avoid double counting, given that the showering process generates additional jets. On the other hand Alpgen lacks the loop corrections that should accompany the higher order tree diagrams. This shows up as more sensitivity to the choice of renormalization scale. For here and in the following we choose $\sqrt{\hat{s}}/2$ for the renormalization scale, which for example approaches $m_t$ near the $t\overline{t}$ threshold. We again choose the CTEQ6M PDF to simplify comparison to MC@NLO, and we ensure that the same PDF is used by Herwig. To run Alpgen we use 4.0 as the maximum pseudorapidity and 0.6 as the minimum jet separation. See \cite{P} for a comparison of Alpgen and MC@NLO and for a description of the MLM matching procedure used by Alpgen.

The high $H_T$ tail of the distribution, of interest for the background, is sensitive to the higher jet multiplicities. We use Alpgen to generate samples of $t\overline{t}+0$, $t\overline{t}+1$ and $t\overline{t}+2$ partons. We note that the lower multiplicity samples become relatively more important for an increasing value of the minimum jet $p_{T}$ parameter used by Alpgen, $p_T^{min}$. We wish to choose $p_T^{min}$ large enough so that the highest jet multiplicity sample does not completely dominate the high $H_T$ tail. (The highest multiplicity sample is inclusive and is sensitive to the parton showering performed by Herwig; we wish to avoid this reliance on Herwig.) We will display results for $p_T^{min}=80$, but in principle results should be fairly independent of the $p_T^{min}$ choice. Indeed we find similar results for  $p_T^{min}=120$ GeV and $p_T^{min}=40$ GeV. In the latter case we have to consider jet multiplicities up to and including the $t\overline{t}+4$ parton sample, again in order for the highest multiplicity sample not to completely dominate the high $H_T$ region.

For the signal we don't need the full Alpgen machinery since we are not on a tail of a distribution in this case. But for a fair comparison of signal and background we nevertheless use Alpgen to produce a $t'\overline{t'}$ sample, although with no extra partons and with jet matching turned off. To model the $t'$ with Alpgen we do the same as with MC@NLO, and simply increase the mass of the $t$ to 600 GeV. An advantage of Alpgen over MC@NLO is that the former incorporates spin correlations in the fully inclusive decays of $t$ and $t'$, although we do not expect this to have much effect on our results.

We find that the Alpgen-Herwig results in Fig.~(4) are strikingly similar to the MC@NLO-Herwig results. Note that we have not included K-factors that would be necessary to bring the total cross sections in line with MC@NLO results. In any case the handling of the higher jet multiplicities in Alpgen, arguably in manner more correct that in MC@NLO, does little to degrade the $t'$ mass reconstruction. If anything these results again suggest that perturbative effects beyond lowest order tend to enhance $S/B$.
\begin{center}\includegraphics[scale=0.28]{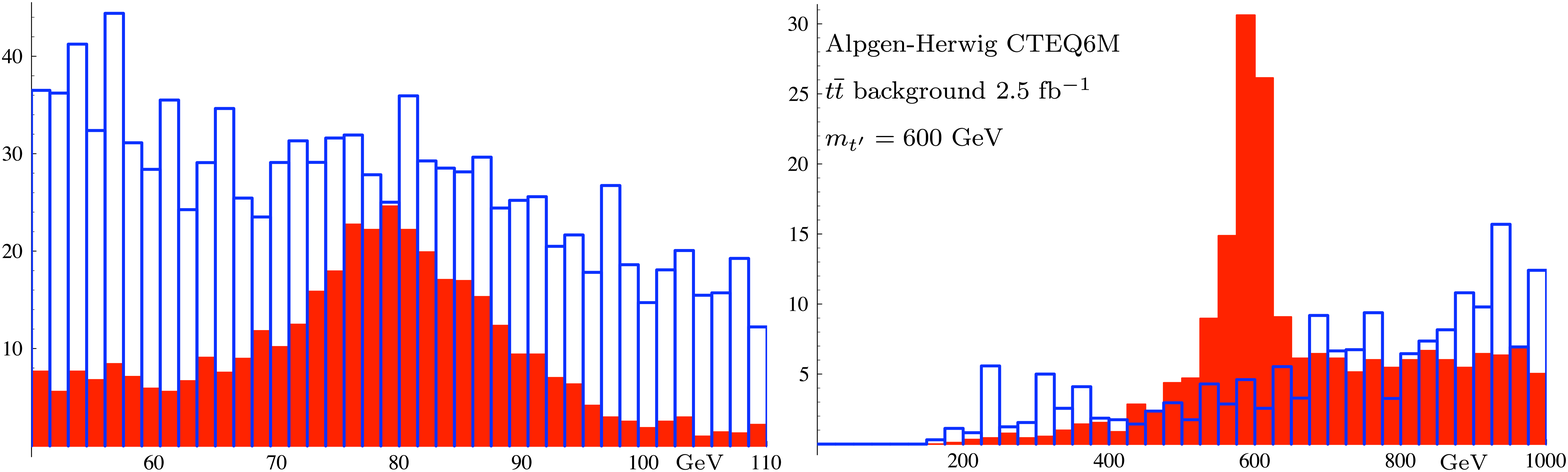}\end{center}
\vspace{-1ex}\noindent Figure 4: Signal versus $t\overline{t}$ background, using Alpgen-Herwig.
\vspace{2ex}

\section{Event generation involving Pythia}
As a further check we would like to use Alpgen in conjunction with Pythia, but first we need to discuss our use of Pythia. In the following Pythia will also be used in stand-alone mode, as well as with Madgraph. We wish Pythia to be used consistently in these three contexts, so that for processes that all three methods can describe, they give the same results. The renormalization scale, used as an argument for parton distributions and for $\alpha_s$ at the hard interaction, is specified explicitly in stand-alone Pythia. Alternatively Alpgen and Madgraph can pass this scale to Pythia on an event-by-event basis. Another important scale in Pythia is the maximum parton virtuality allowed in $Q^2$-ordered space-like showers (initial state radiation). We will refer to this as a phase space cutoff. We find that the high $H_T$ tail of our distributions, important for determining the background, is sensitive to this cutoff. There is much less sensitivity to a corresponding cutoff for time-like showers (final state radiation).

In Pythia, \verb$MSTP(32)=4$ (specifies $\hat{s}$) and \verb$PARP(34)=0.25$ (the pre-factor) gives $\hat{s}/4$ as the square of the renormalization scale. In stand-alone Pythia the phase space cutoffs, space-like and time-like, are determined (when \verb$MSTP(68)=0$) by the factors \verb$PARP(67)$ and \verb$PARP(71)$ that multiply $\hat{s}$ in our case. To get the same cutoffs in Alpgen-Pythia and Madgraph-Pythia, assuming that $\sqrt{\hat{s}}/2$ is the renormalization scale in Alpgen or Madgraph, requires a rescaling \verb$PARP(67)$ $\rightarrow$ \verb$PARP(67)/PARP(34)$ and \verb$PARP(71)$ $\rightarrow$ \verb$PARP(71)/PARP(34)$ in Pythia when using the external events.

We display the Alpgen-Pythia results in Fig.~(5). The same unweighted Alpgen events that were passed through Herwig are passed through Pythia, with the latter again using the CTEQ6M PDF. We find that the Alpgen-Herwig and Alpgen-Pythia results are in excellent agreement.
\begin{center}\includegraphics[scale=0.28]{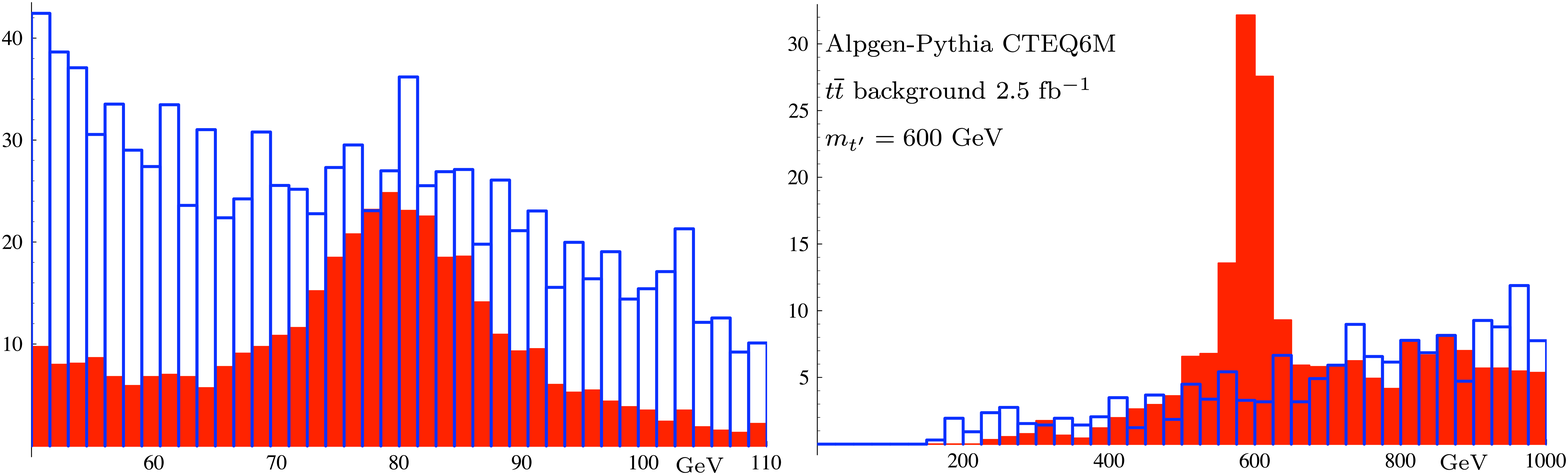}\end{center}
\vspace{-1ex}\noindent Figure 5: Signal versus $t\overline{t}$ background, using Alpgen-Pythia.
\vspace{2ex}

We now consider results of stand-alone Pythia. Pythia has recently provided built in choices for ``tunes'' of the various parameters and options in the modeling of initial and final state radiation and the underlying event, where the latter includes beam remnants and multiple parton interactions.  First we consider the DW tune, developed by R.~D.~Field by testing against CDF data \cite{L}. It has \verb$PARP(67)=2.5$. We will now also switch to the CTEQ5L PDF since this is assumed by the Pythia tunes.\footnote{We also used the DW settings for the Alpgen-Pythia runs, which in that case is not strictly the DW tune due to the different PDF used.}  The results for the DW tune are shown in Fig.~(6), and we see a $S/B$ that is somewhat smaller than previous results of MC@NLO and Alpgen. This reinforces our previous conclusions regarding the role of perturbative corrections.
\begin{center}\includegraphics[scale=0.28]{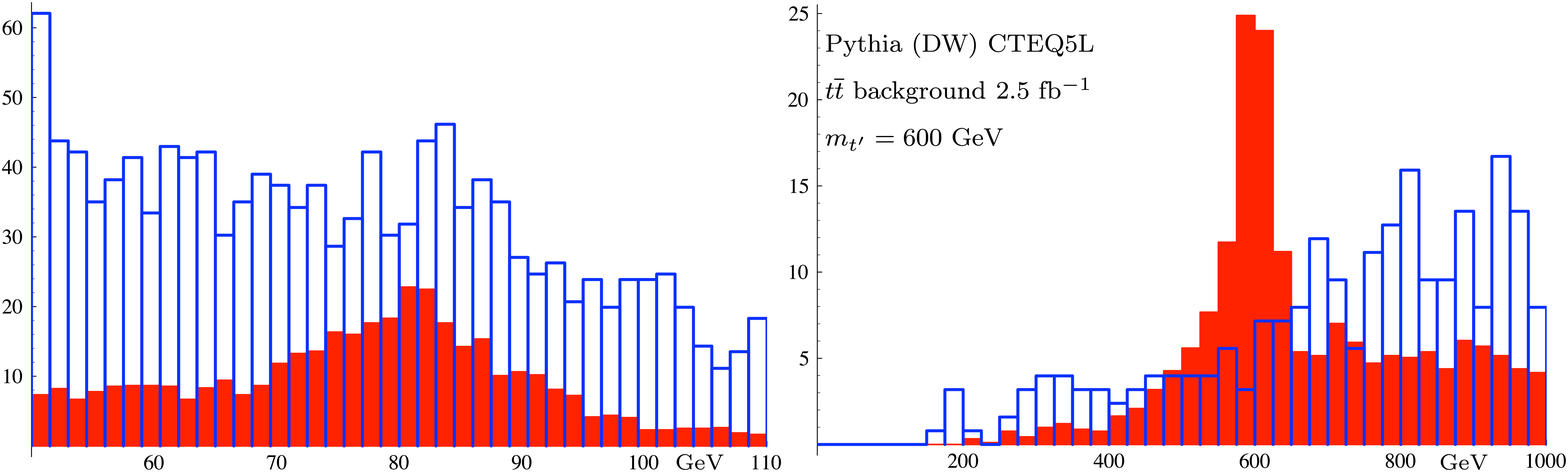}\end{center}
\vspace{-1ex}\noindent Figure 6: Signal versus $t\overline{t}$ background, using stand-alone Pythia with tune DW.
\vspace{2ex} 

It is of interest to consider other tunes involving the different showering and underlying event models available in Pythia.  Tune DW is based on the older $Q^2$-ordered shower model, but there are newer models based on $p_T$-ordered showers. There are four such Sandhoff-Skands tunes \cite{H}, S0, S0A, S1, S2, differing mainly by the color reconnection model they use. Although the value of \verb$PARP(67)$ is not used in these tunes, a low or high phase space cutoff can be chosen with \verb$MSTP(68)=0$ or \verb$3$ respectively. We choose the former and we comment more on this below.  Although the new models are hopefully more realistic than the old, they may not be as well tested or tuned. In addition the new models appear only to work in stand-alone Pythia. Among these we choose to focus on tune S0A, motivated partly by the fact that it shares with tune DW the same value for \verb$PARP(90)=0.25$ (the energy scaling of the infrared cutoff in the underlying event model). Also, its color reconnection model is less computationally intensive than S1 or S2. (Of special note is tune S1 which runs extremely slow and is the only tune to give a lower $S/B$ than tune S0A.) The results for the S0A tune are shown in Fig.~(7). A further drop in $S/B$ from the DW tune is evident.
\begin{center}\includegraphics[scale=0.28]{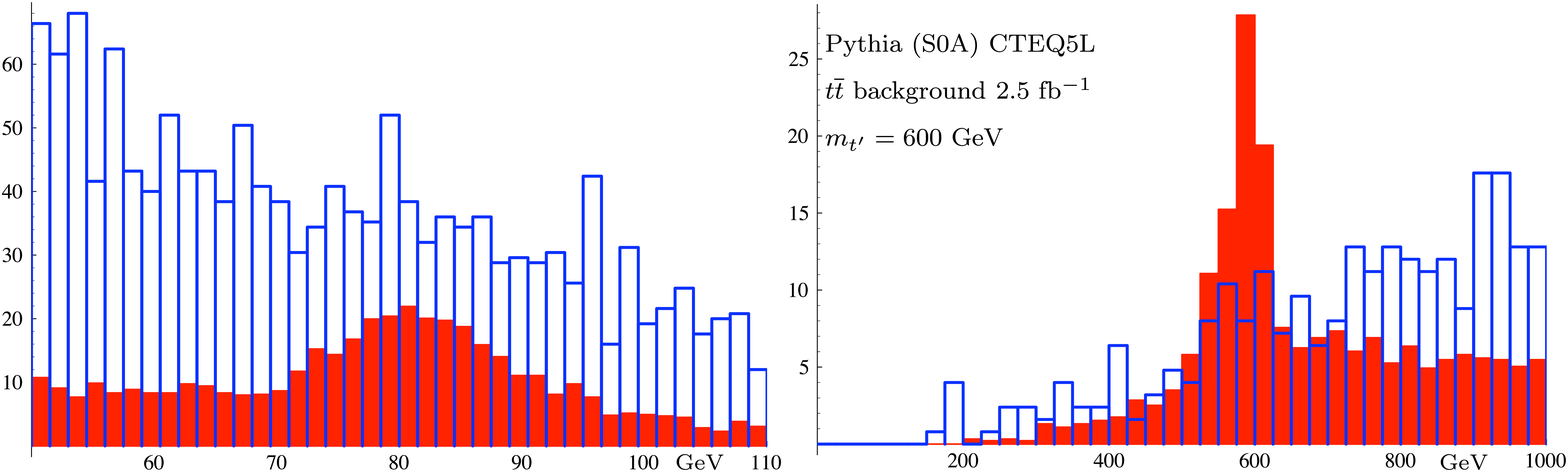}\end{center}
\vspace{-1ex}\noindent Figure 7: Signal versus $t\overline{t}$ background, using stand-alone Pythia with tune S0A.
\vspace{2ex}

In Fig.~(8) we compare the high $H_T$ tails of the various cases. We omit the stand-alone Herwig result already shown in Fig.~(3), which is larger than any here. The largest in Fig.~(8) is from tune S0A and we find that all four Sandhoff-Skands tunes give very similar results for the $H_T$ tail. We see that the relative sizes of these high $H_T$ tails are inversely related to the observed $S/B$ ratios.
\begin{center}\includegraphics[%
  scale=0.8]{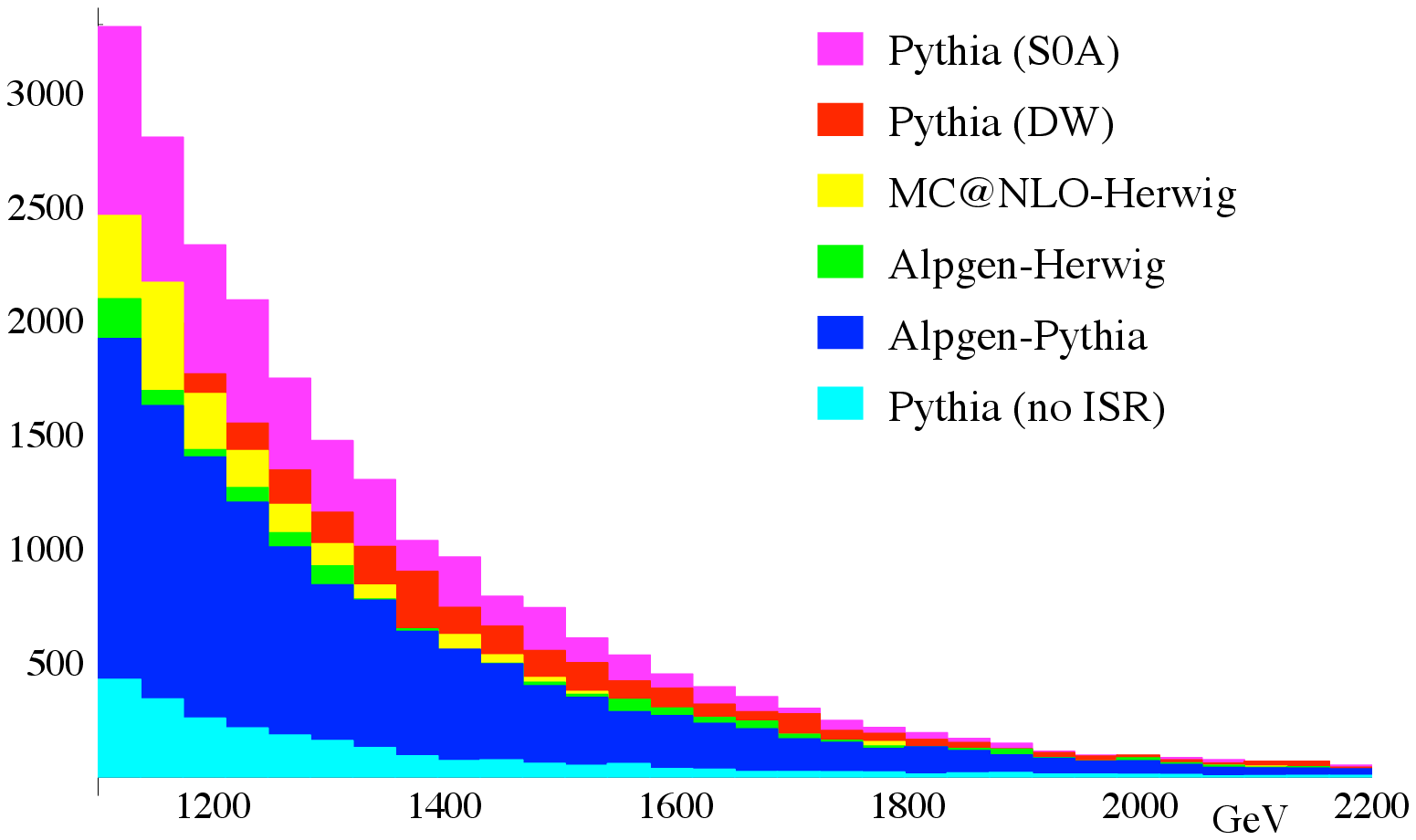}\hspace{-1pt}\end{center}
\vspace{-2ex}
\noindent Figure 8: Distributions on the high $H_T$ tail.
\vspace{2ex}

The distribution much smaller than the others in Fig.~(8) corresponds to the case of turning off initial state radiation in the DW tune of stand-alone Pythia. This last result makes clear just how important the initial state radiation is to the background estimate; in fact the bulk of the background is due to it. Physically this suggests that the high $H_T$ tail of the distribution is receiving significant contributions from scatterings producing $t\overline{t}$ at lower energy, where the cross section is larger, since the remaining energy can come from initial state radiation. Initial state radiation also stimulates multiple parton interactions, which adds to the activity. There has been some discussion \cite{G} of initial state radiation and the motivation to increase the phase space cutoff to better fit $p_T$ distributions of jets. But increasing this cutoff (using the default \verb$MSTP(68)=3$ for the new tunes for example) will significantly inflate the $H_T$ tail further, and thus there appears to be some tension in the attempt by stand-alone Pythia to model simultaneously both the jet $p_T$ and $H_T$ distributions.

\section{Another background and variations}
We now turn to a brief discussion of the $W+\rm{jets}$ background, where the $W$ decays hadronically and at least one jet is mistagged as a $b$-jet.
For this we turn to Madgraph. (In Alpgen only the leptonic $W$ decay is incorporated for this process.) We continue to make the choice of $\sqrt{\hat{s}}/2$ for the renormalization scale (in Madgraph a modification of \verb$setscales.f$ is needed) and CTEQ5L for the PDF. Just from kinematics, for a mistagged $b$-jet and $W$ to have a combined invariant mass in the signal region typically requires that there be at least one other hard jet in the process. We thus focus on the $W+2$ jet process at the partonic level. (We have confirmed that it generates larger background than the $W+1$ jet process). In Madgraph we require a minimum $p_T$ of 120 GeV for each of the two jets (and we confirm that a 100 GeV cut does not increase the background). The size of this background of course depends on the $b$-mistag rate from light partons (mainly gluons), for which we have chosen the quite conservative value of 1/30. The results in Fig.~(9) are again encouraging, with this background being comparable to the $t\overline{t}$ background.
\begin{center}\includegraphics[scale=0.28]{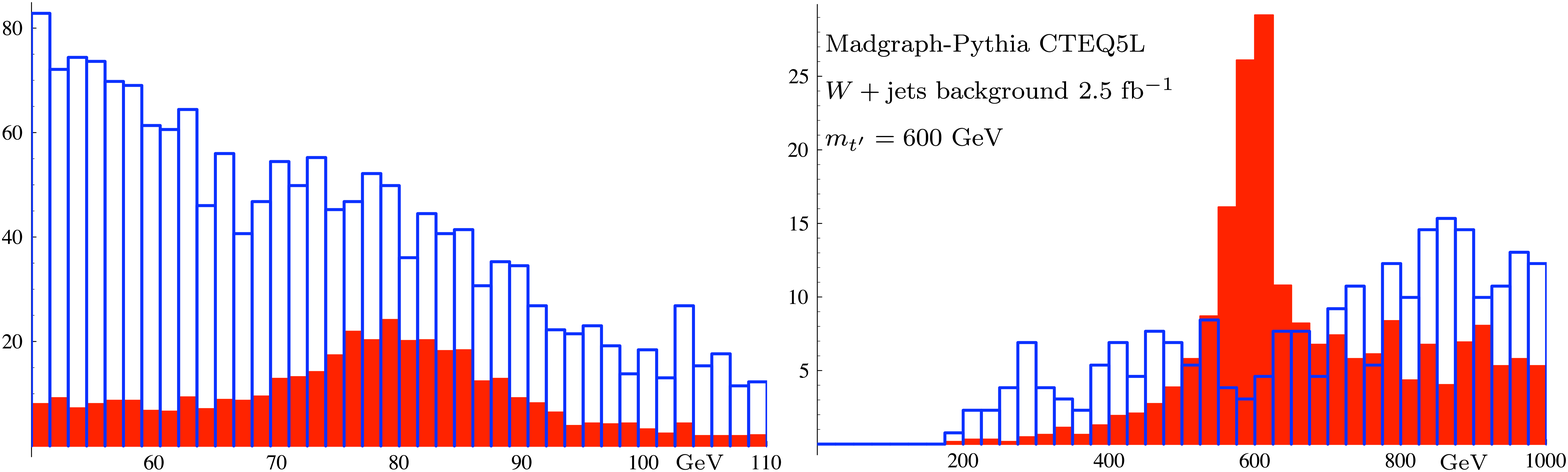}\end{center}
\vspace{-1ex}\noindent Figure 9: Signal versus $W+\rm{jets}$ background, using Madgraph-Pythia.
\vspace{2ex}

More study of the $W+\rm{jets}$ background is warranted, but we note that if necessary it can be significantly reduced relative to the $t\overline{t}$ background. This is done by requiring an additional $b$-tag and/or a lepton and/or missing energy, at the expense of lower statistics. We do not enter into a detailed discussion of other possible backgrounds here, but the ones we have briefly checked, including $b\overline{b}+\rm{jets}$, $(W/Z)b\overline{b}$, $Z+\rm{jets}$, and $(WW/ZZ/WZ)+\rm{jets}$, all appear to be smaller than $W+\rm{jets}$.  More problematic to estimate is the QCD multijet background, and a more detailed study is left for future work. Indeed it may be that this background could force lepton and/or missing energy requirements in our event selection.

Thus far we have considered $m_{t'}=600$ GeV, and so here we briefly consider increasing this mass to 800 GeV. The only difference is that we scale up the $\Lambda_{\rm top5}$  and $\Lambda_b$ cuts by a factor of $4/3$. We display the results for MC@NLO in Fig.~(10) and for the stand-alone Pythia case with the DW tune in Fig.~(11). The comparison is between a case with NLO effects and a case without, and the differences that we have noted before are now accentuated for $m_{t'}=800$ GeV.
\begin{center}\includegraphics[scale=0.28]{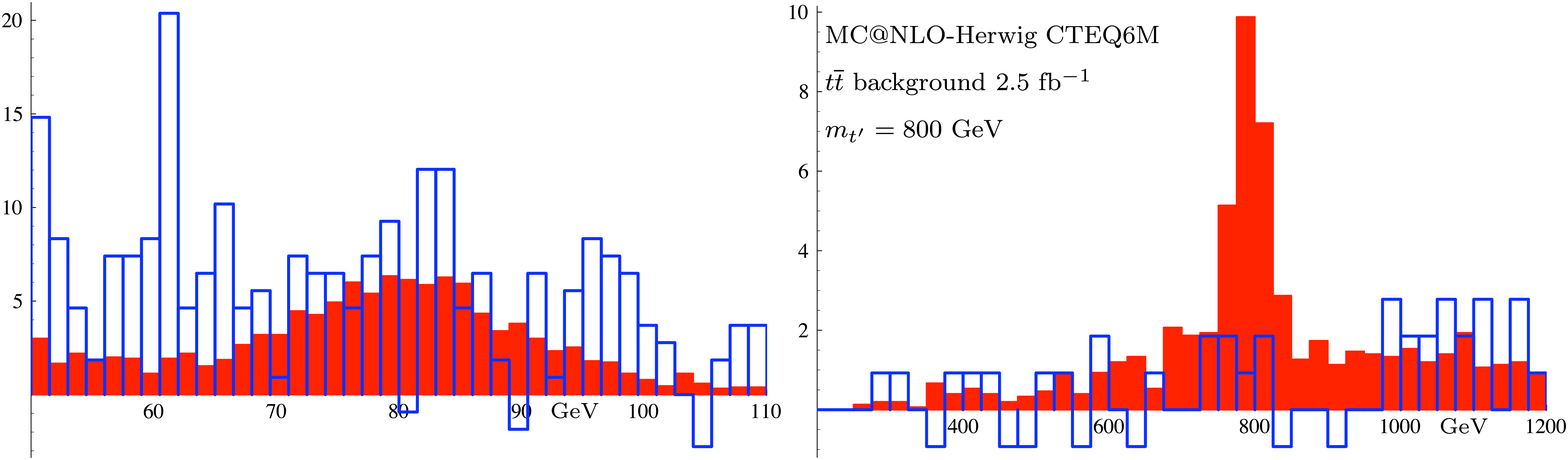}\end{center}
\vspace{-1ex}\noindent Figure 10: Signal versus $t\overline{t}$ background with $m_{t'}=800$ GeV, using MC@NLO-Herwig.
\vspace{2ex}
\begin{center}\includegraphics[scale=0.28]{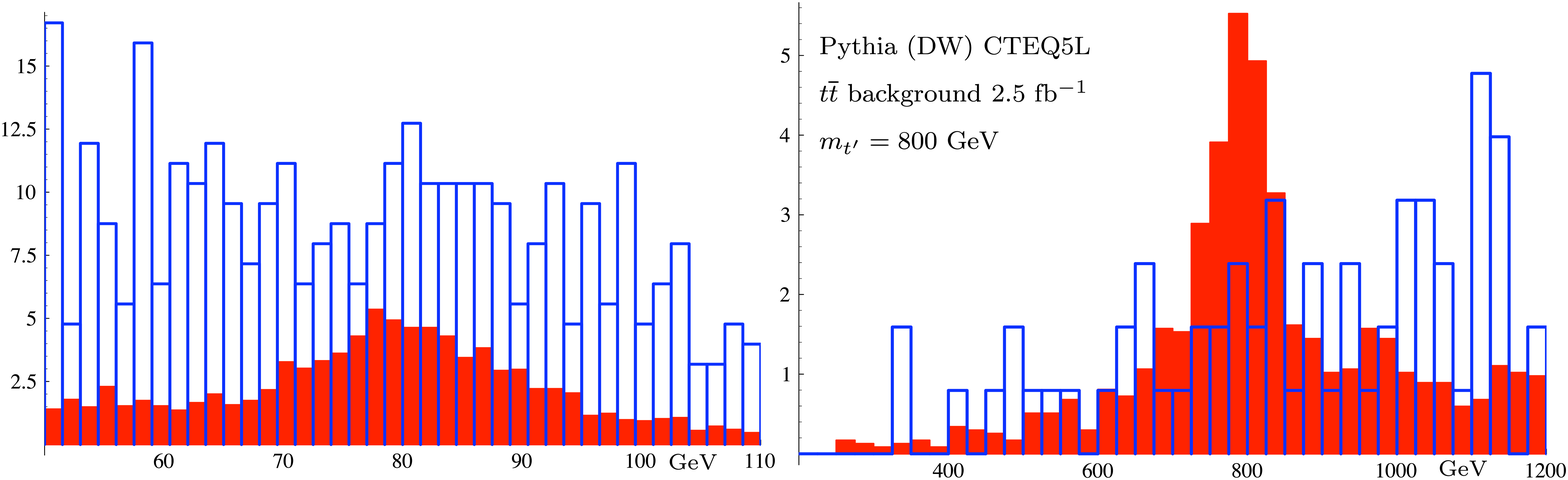}\end{center}
\vspace{-1ex}\noindent Figure 11: Signal versus $t\overline{t}$ background with $m_{t'}=800$ GeV, using Pythia with tune DW.
\vspace{2ex}

Finally, we would like to consider the use of the $k_T$ jet finding algorithm available in PGS4. We keep everything else the same except that we set the parameter analogous to the cone size to 0.5. We display the results for the stand-alone Pythia case with DW tune in Fig.~(12). A comparison to Fig.~(6) illustrates how the $k_T$ jet finder has been unable so far to match the performance of the cone-based jet finder.
\begin{center}\includegraphics[scale=0.28]{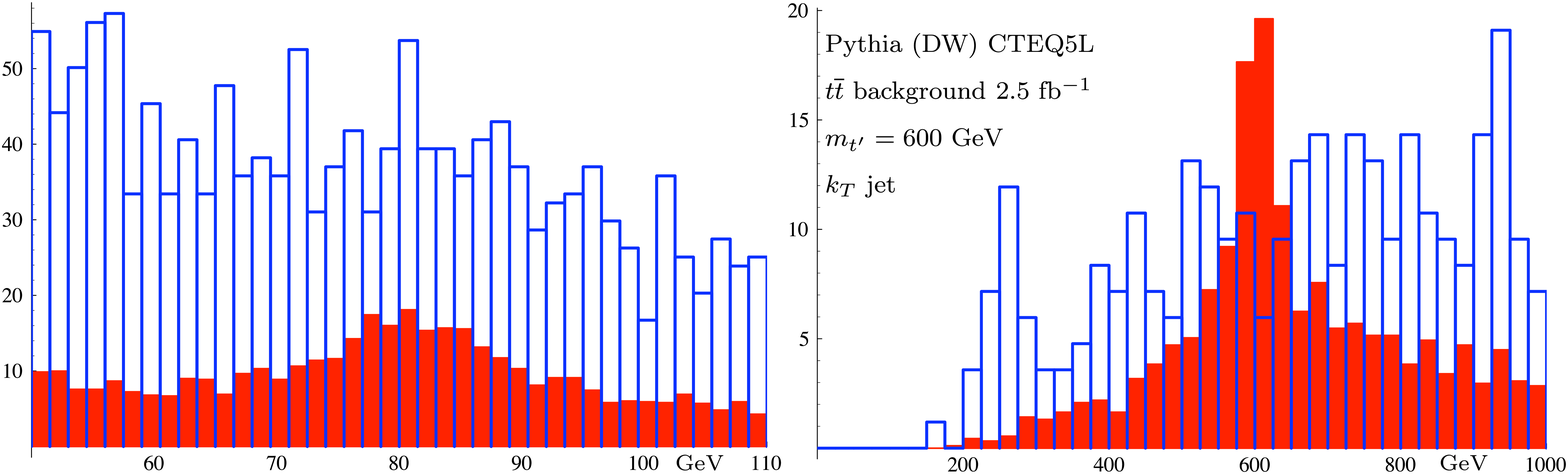}\end{center}
\vspace{-1ex}\noindent Figure 12: Signal versus $t\overline{t}$ background using stand-alone Pythia with tune DW, and using the $k_T$ jet finding algorithm.
\vspace{2ex}

In summary, event selection based on the use of single jet invariant masses and cone-based jet finding provides a very encouraging signal to background ratio for the search for heavy quarks. It appears to survive the various effects that may increase the background estimates, even for just a few inverse femtobarns of data. The background sensitivity to initial state radiation creates one of the larger uncertainties, as does the range of results arising from the various models and tunes available in Pythia. On the other hand the higher order effects as modeled by MC@NLO and Alpgen appear to improve the signal to background ratio. Further analysis and refinements of the method, and the consideration of other applications where $W$ identification is needed, will be left for future work \cite{Q}.

\section{But is a fourth family worth looking for?}

The present theoretical bias against the consideration of a fourth family is basically a reflection of our present lack of understanding of the origin of flavor. Most attention is focussed on the origin of the electroweak symmetry breaking, and a flavor structure is usually simply imposed so as to accommodate the known three families. In this sense flavor physics and the physics of electroweak symmetry breaking have not been well integrated into a common framework. The discovery of a fourth family and the realization that this is very much connected with electroweak symmetry breaking would bring these two issues together. In particular the origin of mass of the light quarks and leptons would become a question of how the heavy masses are fed down to the lighter masses. In the absence of a Higgs scalar and associated Yukawa couplings, one would have to consider effective four-fermion operators as the mechanism for feeding mass down. Since such operators are naturally suppressed by the mass scale of their generation, one is led towards new physics at energy scales not too far above the electroweak scale. A fourth family, and its implication that there is no light Higgs, no low-energy supersymmetry, and no evidence for any required fine-tuning, would shift the focus away from theories of much higher energy scales, and towards the dynamics of a theory of flavor.

\section*{Note added}
Upon the completion of this work the recent work by W. Skiba and D. Tucker-Smith \cite{T} was brought to the author's attention. They also make use of single jet invariant masses, and they reference a further example of earlier work using this technique \cite{U}.

\section*{Acknowledgments}
The author thanks John Conway and Johan Alwall for useful communications. I also thank Brian Beare for his input. This work was supported in part by the National Science and Engineering Research Council of Canada.

\end{document}